\documentstyle[12pt]{article}
\def\omm{1--matrix model}
\def\omms{1--matrix models}
\def\qmm{q--matrix models}
\def\mmm{multi--matrix models}

\def\kph{{KP hierarchy}}
\def\ln{{\rm ln}}

\def\TL{{\tilde L}}
\def\TM{{\tilde M}}
\def\a{\begin{eqnarray}}
\def\b{\end{eqnarray}}
\def\0{\nonumber}
\def\ba{\begin{array}}
\def\ea{\end{array}}
\def\noal{\noalign{\vskip10pt}}

\def\PBC{{\bar{\cal P}}(q)}

\def\q{{\bar{\cal Q}}}

\def\al{{\alpha}}
\def\lm{{\lambda}}

\def\ch{{\cal H}}

\def\cm{{\cal M}}
\def\cq{{\cal Q}}

\def\rtr{{\rm Tr}}
\renewcommand{\theequation}{\thesection.\arabic{equation}}

\newlength{\extraspace}
\newlength{\extraspaces}
\newcounter{dummy}
\newcommand{\ai}{
\addtocounter{equation}{1}
\setcounter{dummy}{\value{equation}}
\setcounter{equation}{0}
\renewcommand{\theequation}{\thesection.\arabic{dummy}\alph{equation}}
\begin{eqnarray}
\addtolength{\abovedisplayskip}{\extraspaces}
\addtolength{\belowdisplayskip}{\extraspaces}
\addtolength{\abovedisplayshortskip}{\extraspace}
\addtolength{\belowdisplayshortskip}{\extraspace}}
\newcommand{\bj}{
\end{eqnarray}
\setcounter{equation}{\value{dummy}}
\renewcommand{\theequation}{\thesection.\arabic{equation}}}
\def\d{{\partial}}
\def\dtl{{\tilde\d}}

\newcommand{\ddlm}[1]{{\partial \over \partial \lm_{#1}}}

\newcommand{\res}{{\rm res}\,}
\begin{document}
\begin{flushright}
SISSA-ISAS 211/92/EP\\
hep-th/9212070
\end{flushright}
\vskip0.5cm
\centerline{\LARGE\bf Multi--matrix models without continuum limit}
\vskip0.3cm
\centerline{\large  L.Bonora}
\centerline{International School for Advanced Studies (SISSA/ISAS)}
\centerline{Via Beirut 2, 34014 Trieste, Italy}
\centerline{INFN, Sezione di Trieste.  }
\vskip0.5cm
\centerline{\large C.S.Xiong\footnote{Address after December 1, 1992: Inst.
Theor.Phys., Academia Sinica, P.O. Box 2735, Beijing 100080, China.}}
\centerline{International School for Advanced Studies (SISSA/ISAS)}
\centerline{Via Beirut 2, 34014 Trieste, Italy}
\vskip5cm
\abstract{We derive the discrete linear systems associated to multi--matrix
models, the corresponding discrete hierarchies and the appropriate coupling
conditions. We also obtain the $W_{1+\infty}$ constraints on the partition
function. We then apply to multi--matrix models the technique,
developed in previous papers, of extracting
hierarchies of differential equations from lattice ones without passing
through a continuum limit. In a q--matrix model we find 2q coupled
differential systems. The corresponding differential hierarchies
are particular versions of the KP hierarchy. We show that the multi--matrix
partition function is a $\tau$--function of these hierarchies.
We discuss a few examples in the dispersionless limit.}

\vfill\eject

\section{Introduction}

In \omms  ~all the information is encoded in the Jacobi matrix $Q$,
a semi--infinite matrix formed by three non--vanishing diagonal lines.
This matrix remains the same in form whatever the polynomial potential is.
This accounts for the relatively simple structure of \omms.
When we turn our
attention to \mmm, the situation drastically changes. First of all we have
several Jacobi matrices characterizing the same model. They are semi--infinite
matrices with a finite or infinite band of non--vanishing diagonal lines.
What is more important, the band size depends on the order of the
polynomial potentials.

As is well--known the purpose of all the matrix model analysis is to extract
flow equations (w.r.t. to the perturbation parameters) for the partition
function, as well as constraints known as string equations.
It is believed that \mmm~  should  give rise in particular to
generalized KdV-hierarchies \cite{Douglas} which in turn relate
to generalized Kontsevich  models \cite{Ko},
\cite{W1},\cite{DW},\cite{W2},\cite{IZ1},\cite{Dijkgraafn}.

As will be shown below, it is not
difficult to extract lattice hierarchies and lattice string equations from
\mmm.
In fact in a q--matrix model we find 2q linear systems and, consequently,
2q lattice hierarchies. They
are generalizations of the Toda lattice hierarchy
and they are sometimes referred to as discrete KP hierarchies.
However significant information can be extracted only from differential
equations. The usual way to obtain a differential hierarchy from
a lattice hierarchy is via a continuum limit.
However, due to the aforementioned features of  \mmm, the application of the
continuum limit technique has not proven as successful as in \omms, and
the research in this subject
has remained at a rather conjectural level. The only exception are the results
obtained in \cite{G},\cite{GN}, concerning the $W$-type constraints satisfied
by \mmm, and the analysis of some particular cases
\cite{Tada},\cite{DEB} (see also \cite{MM}).

In a previous paper \cite{BX1} (see also \cite{BX2}) we introduced a
method to extract a
differential hierarchy from a lattice one. It consists essentially
in using the first flow equation to eliminate the difference operations
in the remaining equations. In this way the latter become differential
equations with respect to the first flow parameter (which is interpreted
as the `space' coordinate).
This procedure can be generalized to the lattice systems specific of
\mmm, and this is what we do in this paper. As our main
result, we associate to any q--matrix model a set of differential
hierarchies of flow equations, together with the appropriate string
equations. The differential hierarchies turn out to be particular versions
(reductions) of the generalized KP hierarchy \cite{X2}. We will refer to these
hierarchies as KP--type hierarchies or, when no confusion is possible,
KP hierarchies for short.
As we will see, the structure of \mmm~  is much richer
than it is usually believed. In particular it contains much more than
the n--th KdV hierarchies.

We recall that the importance of avoiding a continuum limit lies in the
fact \cite{BX2} that the differential hierarchies and string equations so
obtained describe properties of the matrix--model lattice and therefore,
presumably, topological properties.

The matter is arranged as follows. Section 2 is a review of \mmm~
and an occasion to state our notations.
In section 3 we represent an arbitrary multi--matrix model by means of
suitable coupled
discrete linear systems, and emphasize the importance of  the coupling
conditions; the corresponding consistency conditions result in (generalized)
2--dimensional Toda lattice hierarchies and string equations.
In section 4 we work out in detail the $W_{1+\infty}$ constraints
on the partition function for two--matrix models.
In section 5 we pass
from lattice to differential formalism and find the KP structure
of the differential hierarchies pertinent to \mmm.
We also prove that the multi--matrix model partition
function is
in fact the $\tau$ function of the appropriate KP--type hierarchy.
In section 5 we see a few explicit examples of the above differential
hierarchy in the dispersionless limit and prepare the ground for future
developments.

\section{Multi--Matrix Models: Partition Function}

The partition function of the q--matrix model is given by
\a
Z_N(t,c)=\int dM_1dM_2\ldots dM_qe^{TrU}\0
\b
where $M_1,\ldots,M_q$ are Hermitian $N\times N$ matrices and
\a
U=\sum_{\al=1}^qV_{\al}
+\sum_{\al=1}^{q-1}c_{\al,\al+1}M_{\al}M_{\al+1}\0
\b
with potentials
\a
V_{\al}=\sum_{r=1}^{p_{\al}}t_{\al,r}M_{\al}^r\,\qquad \al=1,2\ldots,q.\0
\b
The $p_{\al}$'s are finite or infinite positive integers.

The ordinary procedure to calculate the partition function consists of
three steps \cite{BIZ},\cite{IZ2},\cite{M}:

\noindent
$(i)$. One integrates out the angular parts such that only the
integrations over the eigenvalues are left,
\a
Z_N(t,c)={\rm const}\int\prod_{\al=1}^q\prod_{i=1}^N d\lm_{\al,i}
\Delta(\lm_1)e^{{U}}\Delta(\lm_q),\0
\b
where
\a
{U}=\sum_{\al=1}^q\sum_{i=1}^NV(\lm_{\al,i})+
\sum_{\al=1}^{q-1}\sum_{i=1}^N c_{\al,\al+1}\lm_{\al,i}\lm_{\al+1,i},\0
\b
and $\Delta(\lm_1)$ and $\Delta(\lm_q)$ are Vandermonde determinants.

\noindent
$(ii)$. One introduces the orthogonal polynomials
\a
\xi_n(\lambda_1)=\lambda_1^n+\hbox{lower powers},\qquad\qquad
\eta_n(\lambda_q)=\lambda_q^n+\hbox{lower powers}\0
\b
which satisfy the orthogonality relations
\a
\int  d\lambda_1\ldots d\lambda_q\xi_n(\lambda_1)
e^{V_1(\lm_1)+\mu+V_q(\lm_q)}
\eta_m(\lambda_q)=h_n(t,c)\delta_{nm}\label{orth1}
\b
where
\a
\mu\equiv\sum_{\al=2}^{q-1}\sum_{r=1}^{\infty}t_{\al,r}\lambda_{\al}^r
+\sum_{\al=1}^{q-1}c_{\al,\al+1}\lambda_{\al}\lambda_{\al+1}.\0
\b

\noindent
$(iii)$. We expand the Vandermonde determinants in terms of these orthogonal
polynomials and using the orthogonality relation (\ref{orth1}), we can easily
calculate the partition function
\a
Z_N(t,c)={\it const}~N!\prod_{i=0}^{N-1}h_i\label{parti1}
\b
Like in \omms, knowing the partition function means knowing
the coefficients $h_n(t,c)$'s. Like in \omms, we will show in the next section
that the information concerning the latter can be encoded in suitable
linear systems, which have the advantage of clearly showing the integrable
character of the multi--matrix models.

\centerline{-----------------------}

To end this section let us introduce some convenient notations.
For any matrix $M$, we define
\a
\bigl(\cm\bigl)_{ij}= M_{ij}{{h_j}\over{h_i}},\qquad
{\bar M}_{ij}=M_{ji},\qquad
M_l(j)\equiv M_{j,j-l}.\0
\b
We will say that the element $M_l(j)$ belongs to the $j$--th sector.
As usual we also  introduce a natural gradation
\a
{\rm deg}[E_{ij}]=j-i.\0
\b
For a given matrix $M$, if all its non--zero elements
have degrees in the interval $[a,b]$, then we will simply
write: $M\in [a,b]$.

Moreover $M_+$ will denote the upper triangular part of $M$ (including the main
diagonal), while $M_-=M-M_+$. We will write
\a
{\rm Tr} (M)= \sum_{i=0}^{N-1} M_{ii},\quad\quad\quad \widetilde {\rm Tr}(M)=
\sum_{i=0}^\infty M_{ii}\0
\b

Analogously, if $L$ is a pseudodifferential operator, $L_+$ means the purely
differential part of it, while $L_-=L-L_+$.

\section{The Coupled Discrete Linear Systems in q--Matrix Model}

\setcounter{equation}{0}

In order to extract the discrete linear systems that characterize
our q--matrix model we have to pass through some preliminaries. First
we redefine the orthogonal polynomials in the following way
\a
\Psi_n(\lambda_1)=e^{V_1(\lambda_1)}\xi_n(\lambda_1),
\qquad
\Phi_n(\lambda_q)=e^{V_q(\lambda_q)}\eta_n(\lambda_q).\0
\b
As usual we denote the semi--infinite column vectors with components
$\Psi_0,\Psi_1,\Psi_2,\ldots,$ and  $\Phi_0,\Phi_1,\Phi_2,\ldots,$
by $\Psi$ and $\Phi$, respectively. With these polynomials, the orthogonality
relation (\ref{orth1}) becomes
\a
\int\prod_{\beta=1}^q d\lm_{\beta}\Psi_n(\lambda_1)
e^{\mu}
\Phi_m(\lambda_q)=\delta_{nm}h_n(t,c).\label{orth2}
\b
This orthogonality relation is going to play a crucial role in our analysis.
Next we introduce the following $Q$--type matrices
\a
\int\prod_{\al=1}^q d\lm_{\al}\Psi_n(\lambda_1)
e^{\mu}\lm_{\al}
\Phi_m(\lambda_q)\equiv Q_{nm}(\al)h_m=\q_{mn}(\al)h_n,\quad
\al=1,\ldots,q.\label{Qalpha}
\b
Among them, $Q(1),\q(q)$ are Jacobi matrices: their pure upper triangular
part is $I_+=\sum_i E_{i,i+1}$. Notice also that if we call
\a
Q_{n+1,n}(q)= R_{n+1}\0
\b
we immediately find
\a
h_{n+1}= R_{n+1} h_n\0
\b
Therefore, like in \omms, we can express the partition functions in terms of
$R_n$.

Beside the above $Q$ matrices, we will need two $P$--type matrices, defined by
\a
&&\int\prod_{\al=1}^q d\lm_{\al}\Bigl(\ddlm 1 \Psi_n(\lambda_1)\Bigl)
e^{\mu}\Phi_m(\lambda_q)\equiv P_{nm}(1)h_m\\
&&\int  d\lambda_1\ldots d\lambda_q\Psi_n(\lambda_1)
e^{\mu}\Bigl(\ddlm q \Phi_m(\lambda_q)\Bigl)\equiv P_{mn}(q)h_n
\b

\subsection{The coupling conditions or string equations.}

One important preliminary remark is that
the matrices (\ref{Qalpha}) we
introduced above are not completely independent. More precisely all
the $Q(\alpha)$'s can be expressed in terms of only one of them and
one matrix $P$.
Expressing the trivial fact that the integral of the total derivative of the
integrand in eq.(\ref{orth2}) with respect to $\lm_{\al},1\leq\al\leq q$
vanishes, we can easily derive the constraints or {\it coupling conditions}
\ai
&&P(1)+c_{12}Q(2)=0,\label{coup1}\\
&&c_{\al-1,\al}Q(\al-1)+V'_{\al}+c_{\al,\al+1}Q(\al+1)=0,\qquad
2\leq \al\leq q-1,\label{coup2}\\
&&c_{q-1,q}Q(q-1)+\PBC=0.\label{coup3}
\bj
where we use the notation
\a
V'_{\al}=\sum_{r=1}^{p_{\al}}rt_{\al,r}Q^{r-1}(\al),\qquad \al=1,2,\ldots, q\0
\b
These conditions play an extremely important role in the study of \mmm.
We will also refer to them as {\it string equations}.

Before proceeding further, let us pause
a bit to make a few comments.
\begin{itemize}
\item
It is just these coupling conditions that lead to the famous
$W_{1+\infty}$--constraints on the partition function at the discrete level
\cite{IM}(see section 4).
\item
These conditions explicitly show that the Jacobi matrices depend
on the choice of the potentials.
We can immediately see that
these coupling conditions completely determine
the degrees of the matrices $Q(\al)$.
Since $P(1)-V'_1$ and $P(q)-{\bar V'_q}$ are {\it purely} lower
triangular matrices, a simple calculation shows that
\a
Q(\al)\in[-m_{\al}, n_{\al}],\qquad \al=1,2,\ldots,q\0
\b
where
\a
&&m_1=(p_q-1)\ldots(p_3-1)(p_2-1) \0\\
&&m_{\al}=(p_q-1)(p_{q-1}-1)\ldots(p_{\al+1}-1),\qquad~~~~~2\leq\al\leq q-1\0\\
&&m_q=1 \0
\b
and
\a
&&n_1=1\0\\
&&n_{\al}=(p_{\al-1}-1)\ldots(p_2-1)(p_1-1),
\qquad\qquad~~2\leq\al\leq q-1\0\\
&&n_q=(p_{q-1}-1)\ldots(p_2-1)(p_1-1)\0
\b
These equations show that, if we want a finite band structure for the
$Q(\alpha)$, only a finite number of perturbations is allowed.
Conversely, if we want, say, $Q(1)$ to possess the full discrete KP
structure -- we
should let $p_{\al}\longrightarrow\infty$ for some $\alpha$;
then all these matrices will have infinitely many non--zero
diagonal lines. So, generally speaking, these coupling conditions
reveal the main difference between
multi--matrix models and one--matrix models.
\item
Finally these conditions suggest that there may be a relation between matrix
models and
topological models. As an illustration of this remark, we consider a particular
case -- the two--matrix models. In this situation, the
eqs.(\ref{coup1}--\ref{coup3}) reduce to
\a
P(1)+c_{12}Q(2)=0,\qquad\quad
c_{12}Q(1)+\bar{\cal P}(2)=0,\0
\b
In particular
\a
c_{12}Q_+(2)=-V'_+\bigl(Q(1)\bigl),\qquad
c_{12}\bar\cq_+(1)=-V'_+\bigl(\bar{\cal Q}(2)\bigl)\label{2mconst}
\b
Now, we have $p_1+p_2+1$ series of coordinates, i.e. $p_1+p_2+1$
non--vanishing diagonals of $Q(1)$ and $Q(2)$. The above equations
(\ref{2mconst}) give $(p_1+p_2)$ series of constraints. Therefore
we are only left with one--series of independent coordinates, i.e. one can
express all the elements of $Q(1)$ and $Q(2)$ in terms of the $R_i$'s
(and some coupling constants). This is not surprising since
all the information of the matrix
model is encoded in the partition function, which only involves $R_i$'s.
Finally the remaining coupling conditions
provides more series  of
constraints, so that eventually there are no `local' degrees of freedom left.
This means that, if \mmm~  admit any field theory interpretation,
they should correspond to topological field theories.
\end{itemize}

\subsection{The associated discrete linear systems}.

The derivation of the  linear systems associated to our matrix model
is very simple.  We take the derivatives of eqs.(\ref{orth2})
with respect to the time parameters $t_{\al,r}$, and use
eqs.(\ref{Qalpha}).  We get in this way the time evolution of $\Psi$
and $\Phi$, which can be represented in two different ways:

\vskip0.2cm
\noindent
(*)~~~~{$\underline {Discrete~  Linear~ System~~I}$:}
\a
\left\{\ba{ll}
Q(1)\Psi(\lambda_1)=\lambda_1\Psi(\lambda_1),& \\\noal
{\partial\over{\partial t_{1,k}}}\Psi(\lambda_1)=Q^k_+(1)
\Psi(\lambda_1),&1\leq k\leq p_1,\\\noal
{\partial\over{\partial t_{\al,k}}}\Psi(\lambda_1)=-Q^k_-(\al)
\Psi(\lambda_1),&1\leq k\leq p_{\al};\qquad 2\leq\al\leq q,\\\noal
{\partial\over{\partial\lm}}\Psi(\lambda_1)=P(1)\Psi(\lm_1).&
\ea\right.\label{DLS1}
\b
The corresponding consistency conditions are
\ai
&&[Q(1), ~~P(1)]=1\label{CC11}\\
&&{\partial\over{\partial t_{\al,k}}}Q(1)=[Q(1),~~Q^k_-(\al)]\label{CC12}\\
&&{\partial\over{\partial t_{\al,k}}}P(1)=[P(1),~~Q^k_-(\al)]
\label{CC13}
\bj
(\ref{CC12}) and (\ref{CC13})
are hierarchies, referred to henceforth as discrete KP hierarchies,
whose integrability and meaning will be discussed later on.
The first equation (\ref{CC11}) is often referred to as string equation.
Throughout the paper we
use the latter term also as synonym of coupling conditions (see subsection
3.1).

\vskip0.2cm
\noindent(**)~~~~{$\underline {Discrete~  Linear~ System~~II}$:}
\a
\left\{\ba{ll}
\q(q)\Phi(\lambda_q)=\lambda_q\Phi(\lambda_q),\\\noal
{\partial\over{\partial t_{q,k}}}\Phi(\lambda_q)=\q^k_+(q)
\Phi(\lambda_q),& \\\noal
{\partial\over{\partial t_{\al,k}}}\Phi(\lambda_q)=-\q^k_-(\al)
\Phi(\lambda_q),
&1\leq k\leq p_{\al};\qquad 1\leq\al\leq q-1\\\noal
{\partial\over{\partial\lm_q}}\Phi(\lambda_q)=P(q)\Phi(\lm_q).&
\ea\right.\label{DLS2}
\b
with consistency conditions
\ai
&&[\q(q),~~P(q)]=1,\label{CC21}\\
&&{\partial\over{\partial t_{\al,k}}}\q(q)=[\q(q),~~\q^k_-(\al)]
\label{CC22}\\
&&{\partial\over{\partial t_{\al,k}}}P(q)=[P(q),~~\q^k_-(\al)]\label{CC23}
\bj

\vskip0.2cm
\noindent
(***)~~~~{$\underline {The\, Other\,Discrete~  Linear~ Systems}$:}

One may wonder what are
the equations of motion of the matrices $Q(\al)$'s, with $\alpha \neq 1,q$.
In the light of the coupling conditions it is not surprising to find out
that we have enough information to write them down.
They can actually be extracted from the eqs.(\ref{CC11}--\ref{CC13})
(or (\ref{CC21}--\ref{CC23})) and
eqs.(\ref{coup1}--\ref{coup3}). We have
\ai
&&{\partial\over{\partial t_{\beta,k}}}Q(\al)=[Q^k_+(\beta),
{}~~Q(\al)],\qquad 1\leq \beta\leq\al\label{CC31}\\
&&{\partial\over{\partial t_{\beta,k}}}Q(\al)=[Q(\al),
{}~~Q^k_-(\beta)],\qquad \al\leq \beta\leq q\label{CC32}
\bj
where $k$ runs from $1$ to $p_{\beta}$.

It is tedious, but straightforward to prove that {\it all the flows we have
introduced so far commute with one another.}

It is natural at this point to ask whether eqs.(\ref{CC31},\ref{CC32}) can
be considered as integrability conditions of some linear systems.
The answer is yes, but the relevant linear systems are a bit different
from the ones we met so far. Let us start from the orthogonal
relation (\ref{orth2}) once again, and define two series of functions
\a
\xi^{(\al)}_n(t,\lm_{\al})\equiv\int\prod_{\beta=1}^{\al-1}
d\lm_{\beta}\Psi_n(\lambda_1)e^{\mu_{\al}}.\label{xial}
\b
and
\a
\eta^{(\al)}_n(t,\lm_{\al})\equiv\int\prod_{\beta=\al+1}^{q}
d\lm_{\beta}e^{\nu_{\al}}
\Phi_m(\lambda_q).\label{etaal}
\b
where we denote
\a
&&\mu_{\al}\equiv\sum_{\beta=2}^{\al-1}\sum_{r=1}^{\infty}t_{\beta,r}
\lambda_{\beta}^r
+\sum_{\beta=1}^{\al-1}c_{\beta,\beta+1}\lambda_{\beta}\lambda_{\beta+1}.\0\\
&&\nu_{\al}\equiv\sum_{\beta=\al+1}^{q-1}\sum_{r=1}^{\infty}t_{\beta,r}
\lambda_{\beta}^r+\sum_{\beta=\al}^{q-1}c_{\beta,\beta+1}\lambda_{\beta}
\lambda_{\beta+1}.\0
\b
Obviously we have
\a
\xi^{(1)}_n(t,\lm_{1})=\Psi_n(\lambda_1),\qquad\qquad
\eta^{(q)}_n(t,\lm_{q})=\Phi_m(\lambda_q).\0
\b
but for other values of $\alpha$ one sees immediately that $\xi^{(\al)}$
and $\eta^{(\al)}$ are not polynomials anymore.
However they still satisfy an orthogonality relation
\a
\int d\lm_{\al}\xi^{(\al)}_n(t,\lm_{\al})
e^{V_{\al}(\lm_{\al})}\eta^{(\al)}_n(t,\lm_{\al})
=\delta_{nm}h_n(t,c),\qquad \forall 1\leq\al\leq q.\label{orth3}
\b
The $q$ series  of functions $\xi^{(\al)}_n(t,\lm_{\al})$ span $q$ spaces
$\ch_{\al}$. The $q$ series  of functions
$\eta^{(\al)}_n(t,\lm_{\al})$ also span $q$ spaces $\ch^*_{\al}$'s,
which are dual to $\ch_{\al}$ via
the orthogonal relations (\ref{orth3}) \footnote{
This  structure bears a perhaps not accidental similarity with the structure
of topological field theories\cite{Dijkgraafn}.}.

Now we extract the  spectral equations and the time
evolution of these new functions. From eqs.(\ref{Qalpha}), we immediately
see that
\a
\int d\lm_{\al}\xi^{(\al)}_n(t,\lm_{\al})
e^{V_{\al}(\lm_{\al})}\lm_{\al}\eta^{(\al)}_n(t,\lm_{\al})
=Q_{nm}(\al)h_m(t,c),\qquad  1\leq\al\leq q.\label{qal'}
\b
This tells us that the spectral equations are
\a
&&\lm_{\al}\xi^{(\al)}=Q(\al)\xi^{(\al)},\qquad 1\leq\al\leq q.\label{specal}\\
&&\lm_{\al}\eta^{(\al)}=\q(\al)\eta^{(\al)},\qquad 1\leq\al\leq q.
\label{specal'}
\b
On the other hand, from the definitions (\ref{xial}) and (\ref{etaal}),
and making use of eqs.(\ref{DLS1}) and (\ref{DLS2}), we can derive the
evolution equations of the functions $\xi^{(\al)}$ and $\eta^{(\al)}$,
which take the following form
\a
&&{\partial\over{\partial t_{\beta,r}}}\xi^{(\al)}
=Q^r_+(\beta)\xi^{(\al)},\qquad 1\leq \beta\leq\al-1,\label{eqmxi1}\\
&&{\partial\over{\partial t_{\beta,r}}}\xi^{(\al)}
=-Q^r_-(\beta)\xi^{(\al)},\qquad \al\leq \beta\leq q.\label{eqmxi2}
\b
and
\a
&&{\partial\over{\partial t_{\beta,r}}}\eta^{(\al)}
=\q^r_+(\beta)\eta^{(\al)},\qquad \al+1\leq \beta\leq q,\label{eqmeta1}\\
&&{\partial\over{\partial t_{\beta,r}}}\eta^{(\al)}
=-\q^r_-(\beta)\eta^{(\al)},\qquad 1\leq\beta\leq\al.\label{eqmeta2}
\b

Now equations (\ref{specal}--\ref{eqmeta2}) represent
$2q$ discrete linear systems, two of which coincide with (\ref{DLS1}) and
(\ref{DLS2}), the other $(2q-2)$ being
the additional discrete linear systems which give rise to
the compatibility conditions (\ref{CC31}, \ref{CC32})
and to the appropriate string equations.

Finally we notice that, thanks to the generalized systems
of orthogonal functions just introduced, we can represent
all the linear systems in a unified form as follows
\a
\left\{\ba{ll}
Q(\al)\xi(\lambda_{\al})=\lambda_{\al}\xi(\lambda_{\al}),& \\\noal
{\partial\over{\partial t_{\beta,r}}}\xi(\lambda_{\al})=Q^r_+(\beta)
\xi(\lambda_{\al}),&1\leq \beta<\al,\\\noal
{\partial\over{\partial t_{\beta,r}}}\xi(\lambda_{\al})=-Q^r_-(\al)
\xi(\lambda_{\al}),&\al\leq \beta\leq q.
\ea\right.\label{DLS1al}
\b
and
\a
\left\{\ba{ll}
\q(\al)\eta(\lambda_{\al})=\lambda_{\al}\eta(\lambda_{\al}),& \\\noal
{\partial\over{\partial t_{\beta,r}}}\eta(\lambda_{\al})=\q^r_+(\beta)
\eta(\lambda_{\al}),&\al+1\leq \beta\leq q,\\\noal
{\partial\over{\partial t_{\beta,r}}}\eta(\lambda_{\al})=-\q^r_-(\al)
\eta(\lambda_{\al}),&1\leq \beta\leq\al.
\ea\right.\label{DLS2al}
\b

One could phrase the above results by saying that
introducing
the functions  $\xi^{(\al)}_n(t,\lm_{\al})$'s and
 $\eta^{(\al)}_n(t,\lm_{\al})$ can be considered as a formal reductions of
the  \mmm~  to \omms, since we are left with the orthogonal relations
(\ref{orth3}), which are similar to the ones in \omms. The major difference
comes from the fact that $\xi^{(\al)}_n(t,\lm_{\al})$'s and
$\eta^{(\al)}_n(t,\lm_{\al})$ are very complicated functions rather than
the polynomials \footnote {Does this mean that \mmm~  can be realized
as 1-matrix models if we choose suitable non--polynomial interactions?}.

A comment is in order concerning the plethora of linear systems we found.
First we notice that, due to the coupling conditions (3.7), any $Q(\alpha)$
can be expressed in terms of $Q(1)$ and $P(1)$ or, alternatively, of $Q(q)$
and $P(q)$. This means that the
{\it linear system I} or the {\it linear system II} can be considered
as the fundamental ones. We already noticed that the equations
of motion for any $Q(\alpha)$ can be derived from the flow equations
of these two systems. In other words these equations of motion admit
2q different linearizations.
In the following we will mostly be dealing with the systems $I$ and $II$.
However one cannot exclude that a linear
system different from $I$ and $II$ might be more convenient for
specific purposes.

\subsubsection{Coordinatization of the $Q$ and $P$ matrices}

Henceforth we will use the following coordinatization of the Jacobi matrices
\a
Q(1)=I_++\sum_i \sum_{l=0}^{m_1} a_l(i)E_{i,i-l}, \qquad\qquad\qquad
\q(q)=I_++\sum_i \sum_{l=0}^{m_q} b_l(i)E_{i,i-l}\label{jacobi}
\b
and for the supplementary matrices
\a
Q(\al)=\sum_i\sum_{l=-n_{\al}}^{m_{\al}}T^{(\al)}_l(i)E_{i,i-l},
\quad2\leq\al\leq q-1.
\b
One should never forget that
\a
\left \{ \matrix{ a_l(i)=0 & l>i\cr
                  b_l(i)=0 & l>i\cr
                  T^{(\alpha)}_l(i)=0 &l>i}\right.\label{limit}
\b
Let us start from the coordinatization (\ref{jacobi}). We can see
immediately that
\a
\Bigl(Q_+(1)\Bigl)_{ij}=\delta_{j,i+1}+a_0(i)\delta_{i,j},\qquad
\Bigl(Q_-(q)\Bigl)_{ij}=R_i\delta_{j,i-1}\0
\b
Therefore we can write down the $t_{1,1}$-- and $t_{q,1}$--flows explicitly
\ai
&&{\partial\over{\partial t_{1,1}}}a_l(j)=a_{l+1}(j+1)-a_{l+1}(j)
+a_l(j)\Big(a_0(j)-a_0(j-l)\Big)\label{f11}\\
&&{\partial\over{\partial t_{q,1}}}a_l(j)=R_{j-l+1}a_{l-1}(j)-R_ja_{l-1}(j-1)
\label{fq1}\\
&&{\partial\over{\partial t_{1,1}}}b_l(j)=R_{j-l+1}b_{l-1}(j)-R_jb_{l-1}(j-1)
\label{f11'}\\
&&{\partial\over{\partial t_{q,1}}}b_l(j)=b_{l+1}(j+1)-b_{l+1}(j)
+b_l(j)\Big(b_0(j)-b_0(j-l)\Big)\label{fq1'}\\
&&{\partial\over{\partial t_{1,1}}}T^{(\al)}_l(j)=T^{(\al)}_{l+1}(j+1)-
T^{(\al)}_{l+1}(j)
+T^{(\al)}_l(j)\Big(a_0(j)-a_0(j-l)\Big)\label{fal11}\\
&&{\partial\over{\partial t_{q,1}}}T^{(\al)}_l(j)=R_{j-l+1}T^{(\al)}_{l-1}(j)
-R_jT^{(\al)}_{l-1}(j-1)
\label{falq1}
\bj

We will also need the following coordinatization of $P(1)$ and $P(q)$
\a
P(1)=\sum_i \sum_{l=-\infty}^{s_1}P_l^{(1)}(i)E_{i,i-l}\label{P1coo}\\
P(q)=\sum_i \sum_{l=-\infty}^{s_q}P_l^{(q)}(i)E_{i,i-l}\label{Pqcoo}
\b
where $s_1$ and $s_q$ are suitable positive (possibly infinite) integers.

\subsection{Flow equations for the partition function.}

As a byproduct of the above results we can easily get the
flow equations for the partition function
\a
&&\frac {\d}{\d t_{\al,r}} \ln Z_N(t,c)={\rm Tr}\Bigl(Q^r(\al)\Bigl),
\label{parti2}\\
&&\qquad\qquad  1\leq r\leq p_{\al};\qquad1\leq\al\leq q\0
\b
Using the consistency conditions, we can rewrite these equations in the
following form
\a
&&{\d^2\over{\d t_{1,1}\d t_{\al,r}}}
\ln Z_N(t,c)=\Bigl(Q^r(\al)\Bigl)_{N,N-1},\label{parti3}\\
&&\qquad\qquad  1\leq r\leq p_{\al};\qquad1\leq\al\leq q\0
\b
or equivalently
\a
&&{\d^2\over{\d t_{q,1}\d t_{\al,r}}}
\ln Z_N(t,c)=\Bigl(\q^r(\al)\Bigl)_{N,N-1},\label{parti3'}\\
&&\qquad\qquad  1\leq r\leq p_{\al};\qquad1\leq\al\leq q\0
\b
We end this subsection with one more remark. The  information concerning \mmm~
is encoded in the discrete hierarchy (\ref{CC31}--\ref{CC32}).
Once we  solve these equations, we can
reconstruct the partition functions of the matrix models:
they will automatically satisfy (\ref{parti2}).

\subsection{The hidden 2-dimensional Toda lattice}

In subsection 3.2 we have shown that multi--matrix models are
equivalent to certain discrete linear systems subject to some constraints.
If we ignore the coupling conditions, the discrete linear systems
lead to discrete hierarchies. In fact we will show that these discrete or
lattice hierarchies, in the $q=2$ case, are nothing but
2--dimensional Toda lattice hierarchies \cite{UT}\cite{AS}, and in the
generic $q$ case,
they are 2--dimensional Toda lattices with additional flows.

Let us start from the first flow equations written above.
In particular, from eq.(\ref{f11}), we have
\a
{\partial\over{\partial t_{q,1}}}a_0(j)=R_{j+1}-R_j\label{rfq1}
\b
while, from(\ref{parti2}), one can show that
\a
{\partial\over{\partial t_{1,1}}}R_j=R_j\bigl(a_0(j)-a_0(j-1)\bigl)\label{rf11}
\b
Combining these two equations we obtain the following 2-dimensional Toda
lattice equation
\a
{\partial^2\over{\partial t_{1,1}\partial t_{q,1}}}\ln R_j=
R_{j+1}-2R_j+R_{j-1}\label{toda1}
\b
In fact in terms of the coordinates $\phi_j=\ln h_j$, the above
equation takes the familiar form
\a
{\partial^2\over{\partial t_{1,1}\partial t_{q,1}}}\phi_j=
e^{\phi_{j+1}-\phi_j}-e^{\phi_j-\phi_{j-1}}\label{toda2}
\b
We could have gotten the same result from
\a
{\partial\over{\partial t_{1,1}}}b_0(j)=R_{j+1}-R_j\label{rfq1b}
\b
and from
\a
{\partial\over{\partial
t_{q,1}}}R_j=R_j\bigl(b_0(j)-b_0(j-1)\bigl)\label{rf11b}
\b
the latter being obtained again from (\ref{parti2}).

(\ref{toda2}) is nothing but the two--dimensional Toda lattice equation.
In the $q=2$ case (i.e. 2--matrix model) the other time flows generate
the complete 2--dimensional Toda lattice hierarchy \cite{UT}.
{}From the analysis in subsection 3.2, we know that the system
is restricted by the coupling conditions, therefore
we can say that 2--matrix models are nothing but constrained 2--dimensional
Toda lattices.
The \qmm~  with $q>2$
contains not only the Toda lattice hierarchy, but also additional
series of flows.

It is perhaps interesting to remark that,
from eqs.(\ref{toda1}) and (\ref{toda2}), the parameters
$t_{1,1}$ and $t_{q,1}$ could be interpreted as
space coordinates while the other parameters are really time flow parameters.
Therefore the \mmm~  hierarchies could underlie a 2+1 D theory.

\subsection{Integrability}

The discrete linear systems ensuing from \mmm~  are integrable as is guaranteed
by the fact that all the flows commute with one another. Most integrable
systems
we know possess a bi--hamiltonian structure with Hamiltonians in involutions.
For the linear systems of \mmm~ we are discussing here we can define a
bi--hamiltonian structure, i.e. two compatible Poisson brackets.

Let us be more explicit. Following \cite{X1}, let us consider {\it system I}
and forget, for the time being, the coupling conditions. We simplify the
notation by writing $Q$ instead of $Q(1)$. Bearing in mind the coordinatization
(\ref{jacobi}), we notice that all the functions $a_l(j)$ can be represented
as
\a
f_X(Q)\equiv {\rm Tr} (QX)\equiv <QX>\0
\b
by means of a suitable upper triangular matrix $X$. We introduce two Poisson
brackets as follows
\a
\{ f_X, f_Y\}_1(Q)&=& <Q[X,Y]>\label{br1}\\
\{ f_X, f_Y\}_2(Q)&=& <(XQ)_+YQ>- <(QX)_+QY>\label{br2}\\
&&+<Q{\cal D}I_+Y>- <QYI_+{\cal  D}>\0
\b
where
\a
{\cal D}= \sum_{j=0}^\infty\sum_{l=0}^j [X,Q]_{ll}E_{j+1,j}\0
\b
In terms of the coordinates we have
\a
\{a_m(i), a_n(j)\}_1&=&a_{m+n}(j) \delta_{i,j-n}  -a_{m+n}(i) \delta_{j,i-m}\0
\b
and
\a
\{a_m(i), a_n(j)\}_2&=&a_{n+m+1}(j) \delta_{i,j-n-1}-a_{n+m+1}(i)
\delta_{i,j+m+1}
\0\\
&&+\sum_{l=0}^m \Big(a_l(i) a_{n+m-l}(j)\delta_{i,j-n+l}- a_{n+m-l}(i)a_l(j)
\delta_{i,j+m-l}\Big)\0\\
&&+a_m(i) a_n(j) \sum_{l=j-n}^j (\delta_{il}-\delta_{i-m,l})\0
\b
Introducing the Hamiltonians
$H_k\equiv H_k(1)\equiv \frac {1}{k} {\widetilde{\rm Tr}}(Q^k)$, one can verify
that the two Poisson brackets are compatible, i.e.
\a
\{H_{k+1},f_X\}_1(Q)=\{H_k,f_X\}_2(Q),\0
\b
that they are in involution and that they generate the flows with respect to
$t_{1,k}$.

With minor modifications we could do the same for {\it system II}.
For example, in the 2--matrix model we can find another couple of
compatible Poisson brackets and another set of Hamiltonians $H_k(2)$ in
involutions. As we already noticed the flows generated by the two series
of Hamiltonians commute.
However we are not able to decide whether the $H_k(1)$'s commute with the
$H_k(2)$'s.
If not, this would be an example of integrable systems
with non--commuting hamiltonians.

\section{$W_{1+\infty}$ Constraints in Two--Matrix Models}

\setcounter{equation}{0}

It was shown in \cite{IM} that the path integral of \mmm~  has a large symmetry
that can be translated into algebraic constraints on the partition function.
It is instructive to see how this large symmetry is inherited by the
corresponding linear systems. We will limit ourselves here to two--matrix
models.
Starting from the string equations and the discrete
hierarchies we found in the previous section, we will prove that the partition
function of the two--matrix model satisfies $W_{1+\infty}$--constraints.
To this end it is actually convenient to consider a much more general
model than the one studied so far, which is
characterized by
the most general polynomial coupling between the two matrices. The reason
is that this enlarged model possesses a larger symmetry than the original one.
By using the ensuing algebraic structure our goal will be reached in a much
easier way. Eventually, by reduction,  we will find the result appropriate
to the original 2--matrix model.

\subsection{The generalized coupling conditions}

Let us introduce a more
general polynomial interaction term into the partition function
\a
Z_N(t,C)&=&{\rm const} \int \prod_{\alpha=1}^2 \prod_{i=1}^N
d\lambda_{\alpha,i}
\Delta (\lambda_1)\Delta(\lambda_2)
e^{\cal U},\label{c1a}\\
{\cal U}&=&\sum_{k=1}^{\infty} t_{1,k}\sum_{i=1}^N
\lambda_{1,i}^k+
\sum_{r=1}^{\infty} t_{2,r}\sum_{i=1}^N \lambda_{2,i}^r
+\sum_{a,b\geq 1}C_{ab}\sum_{i=1}^N \lambda_{1,i}^a
\lambda_{2,i}^b\label{c1b}
\b
In order to recover our original two--matrix model we have to simply
set $C_{ab}=0$ except for $C_{11}=c_{12}$, where the latter is the
coupling of section 2 \footnote{At times, eqs.(4.19) and (4.25), we will
use a compact notation in which $C_{0,b}\equiv t_{2,b}$ and $C_{a,0}\equiv
t_{1,a}$}.

We follow the same procedure we followed in section 2 after
integrating out the angular part, and introduce the orthogonal
polynomials
\a
\xi_n(\lm_1,t)=\lm_1^n+\cdots;\qquad
\eta_n(\lm_2,t)=\lm_2^n+\cdots.\0
\b
which satisfy the improved orthogonal relation
\a
\int\,d\lm_1 d\lm_2 \xi_n(\lm_1,t)e^{U}\eta_m(\lm_2,t)
=h_n\delta_{nm}.\label{corth}
\b
where
\a
&&U(\lambda_1,\lambda_2)
=V_1(\lm_1)+V_2(\lm_2)+\sum_{a,b\geq 1}C_{ab}\lm_1^a \lm_2^b,\0\\
&&V_1(\lm_1)=\sum_{k=1}^{\infty} t_{1,k}\lm_1^k,\qquad\qquad
V_2(\lm_2)=\sum_{r=1}^{\infty} t_{2,r}\lm_2^r.\0
\b
Next we define the $Q$ matrices
\a
\int\,d\lm_1 d\lm_2 \xi_n(\lm_1,t)e^{U}\lm_{\al}\eta_m(\lm_2,t)
=Q_{nm}(\al)h_m=\q_{mn}(\alpha) h_n, \quad\quad \alpha=1,2\label{cqal}
\b
and their conjugates
\a
P_\circ (1)\xi=\ddlm 1 \xi,\qquad P_\circ(2)\eta=\ddlm 2 \eta.\0
\b
We appended the label $\circ$ in order to distinguish these operators from
the previously defined $P(1)$ and $P(2)$.

As in section 3 we can get the spectral equations
\a
\lm_1\xi=Q(1)\xi,\qquad \lm_2\eta=\q(2)\eta,\label{cspec}
\b
and the equations of motion of the polynomials, in the form of
two coupled linear systems, which coincide with the {\it systems I and II}
of the previous section when $q=2$.
The only differences can be found when the
coupling conditions are involved.
Noting the  spectral equations, one can easily see that the coupling
conditions become\footnote{Hereafter we will use the notations
\a
&&V_1^{'}(1)=\sum_{k=1}^{\infty} kt_{1,k}Q^{k-1}(1),\0\\
&&V_1^{''}(1)=\sum_{k=2}^{\infty} k(k-1)t_{1,k}Q^{k-2}(1),\0
\b
etc.}
\ai
&&P_\circ(1)+V_1^{'}(1)+\sum_{a,b\geq 1}aC_{ab}Q^{a-1}(1)Q^b(2)=0,\label{c2a}\\
&&\bar {\cal P}_\circ(2)+V_2^{'}(2)
+\sum_{a,b\geq 1}bC_{ab}Q^a(1)Q^{b-1}(2)=0.\label{c2b}
\bj
The time evolution of the partition function is given by eq.(\ref{parti2}),
with $\alpha=1$ and $2$. We can also write down
the coupling dependence of the partition function
\a
{\partial\over{\partial C_{ab}}}\ln Z_N(t,C)={{\rm Tr}}\Bigl(
Q^a(1)Q^b(2)\Bigl),\qquad\qquad \forall a,b\geq 1.\label{c3}
\b
This is all we need for the derivation of the constraints.

\subsection{Virasoro constraints}

The $W_{1+\infty}$--algebraic constraints contain a particularly important
subset, the Virasoro constraints. We begin with them.
Since the discussion is valid simultaneously for both linear systems,
we temporarily omit the system indices and consider a general Jacobi
matrix $Q$ and its conjugate $P$
\a
Q_{ij}=\delta_{j,i+1}+S_j\delta_{i,j}
+L_j\delta_{j,i-1}+\ldots,\qquad\qquad
[Q, P]=1.\label{c4}
\b
As was shown in \cite{BMX} the string equation completely determines
the matrix $P$.
Proceeding as in \cite{BMX}, we can express the string equation as a series
of equations
\a
\rtr\bigg(Q^{n+1}(1)\Bigl(P_\circ(1)+V^{'}(1)+
)+\sum_{a,b\geq 1}aC_{ab}Q^{a-1}(1)Q^b(2)\Bigl)\bigg)=0,~~n\geq-1
\b
Now using the relation between partition function and $Q$ matrix,
these are equivalent to
\a
\bigl({\cal L}^{[1]}_n(1)+T^{[1]}_n(1)\bigl)Z_N(t;C)=0,
\qquad\qquad n\geq-1.
\label{c5}
\b
where
\a
{\cal L}^{[1]}_{-1}(1)&=&\sum_{k=2}^{\infty}kt_{1,k}
{\partial\over{\partial t_{1,k-1}}}+Nt_{1,1},\0\\
{\cal L}^{[1]}_0(1)&=&\sum_{k=1}^{\infty}kt_{1,k}
{\partial\over{\partial t_{1,k}}}+{1\over2}N(N+1),\label{c6}\\
{\cal L}^{[1]}_n(1)&\equiv& \sum_{k=1}^{\infty}kt_{1,k}
{\partial\over{\partial t_{1,k+n}}}
+(N+{{n+1}\over2}){\partial\over{\partial t_{1,n}}}
+{1\over2}\sum_{k=1}^{n-1}{{\partial^2}\over{\partial t_{1,k}
\partial t_{1,n-k}}},~~ n\geq 1.\0
\b
and
\ai
T^{[1]}_{-1}(1)&\equiv&
\sum_{a\geq 2,b\geq 1}aC_{ab}
{\partial\over{\partial C_{a-1,b}}}
+\sum_{b\geq 1}C_{1b}
{\partial\over{\partial t_{2,b}}},\label{c7a}\\
T^{[1]}_n(1)&\equiv&\sum_{a,b\geq 1}aC_{ab}
{\partial\over{\partial C_{a+n,b}}},\qquad n\geq 0,\label{c7b}
\bj
These operators satisfy the Virasoro algebras
\ai
&&[{\cal L}^{[1]}_n(1), {\cal L}^{[1]}_m(1)]
 =(n-m){\cal L}^{[1]}_{n+m}(1),\qquad n,m\geq-1,\label{c8a}\\
&&[{\cal L}^{[1]}_n(1), {\cal T}^{[1]}_m(1)]
 =0,\qquad \qquad \qquad \qquad n,m\geq-1,\label{c8b}\\
&&[T^{[1]}_n(1), T^{[1]}_m(1)]=(n-m)T^{[1]}_{n+m}(1),
\qquad n,m\geq-1,\label{c8c}
\bj
In order to recover the result we are interested in for our
original two--matrix model we should simply
set, in eqs.(\ref{c5}, \ref{c7a}, \ref{c7b}), $C_{ab}=0$ except for
$C_{11}=c_{12}$, where the latter is the
coupling of section 2.

\subsection{Higher rank constraints}

In the previous section we obtained the Virasoro constraints, which may be
referred to as rank $2$ tensorial constraints. In the following
we deal with higher rank constraints.
In order to get the spin--$3$ operators, we introduce the following notations
\a
{\cal V}&\equiv&\sum_{a,b\geq1}C_{ab}Q^a(1)Q^b(2),\0\\
{\cal V}^{'}&\equiv&\sum_{a,b\geq 1}aC_{ab}Q^{a-1}(1)Q^b(2),\0\\
:{\cal V}^2:&\equiv&\sum_{a,b,a^{'},b^{'}\geq 1}C_{ab}C_{a^{'}b^{'}}
Q^{a+a^{'}}(1)Q^{b+b^{'}}(2).\0
\b
{}From the trivial relation
\a
\int\,d\lm_1 d\lm_2 {\d^2\over{\d\lm_1^2}}\Bigl(
\xi_n(\lm_1,t)e^{U}\eta_m(\lm_2,t)\Bigl)=0,\0
\b
we get the following identity
\a
-P_\circ^2(1)+V^{''}(1)+{V^{'}}^2(1)+2V^{'}(1){\cal V}^{'}+:{{\cal V}^{'}}^2:
+{\cal V}^{''}=0,
\label{c9}
\b
We claim that multiplying (\ref{c9}) by $Q^{n+2}(1)$ from the left and
taking the trace, we obtain the rank--$3$ constraints
\a
\bigl({\cal L}^{[2]}_n(1)-T^{[2]}_n(1)\bigl)Z_N(t;C)=0,\qquad
n\geq -2.
\label{c10}
\b
with the definitions
\ai
{\cal L}^{[2]}_{-2}(1)&\equiv&\sum_{l_1+l_2=3}^{\infty}l_1l_2t_{1,l_1}t_{1,l_2}
{\partial\over{\partial t_{1,l_1+l_2-2}}}
+\sum_{l=3}^{\infty}lt_{1,l}\sum_{k=1}^{l-3}
{{\partial^2}\over{\partial t_{1,k}\partial t_{1,l-k-2}}}\0\\
&~&+2N\bigl(\sum_{l=3}lt_{1,l}{\partial\over{\partial t_{1,l-2}}}
+Nt_{1,2}+{1\over2}t^2_{1,1}\bigl),\label{c11a}\\
{\cal L}^{[2]}_{-1}(1)&\equiv&\sum_{l_1,l_2=1}^{\infty}l_1l_2t_{1,l_1}t_{1,l_2}
{\partial\over{\partial t_{1,l_1+l_2-1}}}
+\sum_{l=3}^{\infty}lt_{1,l}\sum_{k=1}^{l-2}
{{\partial^2}\over{\partial t_{1,k}\partial t_{1,l-k-1}}}\0\\
&~&+2N\sum_{l=2}lt_{1,l}{\partial\over{\partial t_{1,l-1}}}
+{\cal L}^{[1]}_{-1}(1)+N^2t_{1,1},\label{c11b}\\
{\cal L}^{[2]}_0(1)&\equiv&\sum_{l_1,l_2=1}^{\infty}l_1l_2t_{1,l_1}t_{1,l_2}
{\partial\over{\partial t_{1,l_1+l_2}}}
+\sum_{l=2}^{\infty}lt_{1,l}\sum_{k=1}^{l-1}
{{\partial^2}\over{\partial t_{1,k}\partial t_{1,l-k}}}\cr
&~&+2(N+1)\sum_{l=1}lt_{1,l}{\partial\over{\partial t_{1,l}}}
+{1\over3}N(N+1)(N+2),\label{c11c}\\
{\cal L}^{[2]}_n(1)&\equiv&\sum_{l_1,l_2=1}^{\infty}l_1l_2t_{1,l_1}t_{1,l_2}
{\partial\over{\partial t_{1,l_1+l_2+n}}}
+\sum_{l=1}^{\infty}lt_{1,l}\sum_{k=1}^{l+n-1}
{{\partial^2}\over{\partial t_{1,k}\partial t_{1,l+n-k}}}\0\\
&~&+{1\over{2(n+3)}}\sum_{l=1}^{n-2}\sum_{k=1}^{n-l-1}(n-l+2)
{{\partial^3}\over{\partial t_{1,l}\partial t_{1,k}\partial t_{1,n-l-k}}}\0\\
&~&+\bigl(N^2+(n+1)N+{1\over6}(n+1)(n+2)\bigl)
{\partial\over{\partial t_{1,n}}}\0\\
&~&+(2N+n+2){\cal L}^{[1]}_n(1)
\qquad\qquad n\geq 1.\label{c11d}
\bj
and
\ai
T^{[2]}_{-2}(1)&\equiv&
\sum_{a+a^{'}\geq3,b,b^{'}\geq 1}aa^{'}
C_{ab}C_{a^{'}b^{'}}{\partial\over{\partial C_{a+a^{'}-2,b+b^{'}}}}\0\\
&&+\sum_{a\geq3,b\geq 1}a(a-1)C_{ab}
{\partial\over{\partial C_{a-2,b}}}\0\\
&&+\sum_{b,b^{'}\geq 1}
C_{1b}C_{1b^{'}}{\partial\over{\partial t_{2,{b+b^{'}}}}}
+2\sum_{b\geq 1}C_{2b}
{\partial\over{\partial t_{2,b}}},\label{c12a}\\
T^{[2]}_n(1)&\equiv&\sum_{a,b,a^{'},b^{'}\geq 1}aa^{'}C_{ab}C_{a^{'}b^{'}}
{\partial\over{\partial C_{a+a^{'}+n,b+b^{'}}}}\0\\
&&+\sum_{a,b\geq 1}a(a-1)C_{ab}
{\partial\over{\partial C_{a+n,b}}},\qquad n\geq -1.\label{c12b}
\bj
They satisfy the following algebra
\ai
\relax[{\cal L}^{[2]}_n(1), {\cal L}^{[1]}_m(1)]
 =(n-2m){\cal L}^{[2]}_{n+m}(1)
 +m(m+1){\cal L}^{[1]}_{n+m}(1)\label{c13a}\\
\relax[{\cal L}^{[2]}_n(1),{\cal L}^{[2]}_m(1)]=
 2(n-m){\cal L}^{[3]}_{n+m}(1)-(n-m)(n+m+3){\cal L}^{[2]}_{n+m}(1)
 \label{c13b}\\
\relax[T^{[2]}_n(1), T^{[1]}_m(1)]=(n-2m)T^{[2]}_{n+m}(1)
 -m(m+1)T^{[1]}_{n+m}(1)\label{c13c}\\
\relax[T^{[2]}_n(1), T^{[2]}_m(1)]=2(n-m)T^{[3]}_{n+m}(1)
 +(n-m)(n+m+3){\cal T}^{[2]}_{n+m}(1)\label{c13d}
\bj
where we have introduced the spin--$4$ operator
\a
T^{[3]}_n(1)&\equiv&\sum_{a_1,b_1,a_2,b_2,a_3,b_3\geq 1}
a_1a_2a_3C_{a_1b_1}C_{a_2b_2}C_{a_3b_3}
{\partial\over{\partial C_{a_1+a_2+a_3+n,b_1+b_2+b_3}}}\0\\
&&+\sum_{a_1,b_1,a_2,b_2\geq 1}{3\over2}a_1a_2\bigl(a_1+a_2
-2\bigl)C_{a_1b_1}C_{a_2b_2}
{\partial\over{\partial C_{a_1+a_2+n,b_1+b_2}}}\0\\
&&+\sum_{a,b\geq 1}a(a-1)\bigl(a-2\bigl)C_{ab}
{\partial\over{\partial C_{a+n,b}}},\qquad n\geq -3.\label{c14}
\b
As for ${\cal L}^{[3]}_n(1)$, it is a very complicated expression, which
will not be written down here. One can use eq.(\ref{c13b})
as its definition.

Now let us prove our claim---the constraints (\ref{c10}). At first,
we notice that
\a
\rtr\Bigl( P_\circ^2(1)\Bigl)=0,\0
\b
So taking the trace of eq.(\ref{c9}), we can easily rewrite it as follows
\a
\bigl({\cal L}^{[2]}_{-2}(1)-T^{[2]}_{-2}(1)\bigl)Z_N(t;C)=0,\qquad
\label{c10'}
\b
which is a particular case of eqs.(\ref{c10}), when $n=-2$.
Then, the algebras (\ref{c13a}) and (\ref{c13c}) guarantee that
all the other constraints in eqs.(\ref{c10}) are true as well.
On the other hand from (\ref{c13b}) and (\ref{c13d}), together with
(\ref{c10}),
we obtain the rank 4 operator constraints,
\a
\Bigl({\cal L}^{[3]}_n(1)+T^{[3]}_n(1)\Bigl)Z_N(t;C)=0,\qquad n\geq-3.
\label{rank4}
\b
We see that
a larger algebraic structure helps us simplifying the calculations drastically.
Therefore it is evident that we should keep introducing higher and higher
tensor operators in order to close the algebra generated by
$\{\bigl({\cal L}^{[1]}_n(1)+T^{[1]}_n(1)\bigl),
n\geq-1; \bigl({\cal L}^{[2]}_n(1)-T^{[2]}_n(1)\bigl), n\geq-2\}$.
For example, from eqs.(\ref{c13b}) and (\ref{c13d}),
we obtain the rank--$4$ operators
${\cal L}^{[3]}_n(1)$ and $T^{[3]}_n(1)$. In turn
the commutators of these operators with ${\cal L}^{[2]}_n(1)$ and
$T^{[2]}_n(1)$ generate the rank--$5$ operators and so on.
In this way we must introduce the whole sequence of operators
\a
T^{[r]}_n(1)&\equiv&\sum_{a_1,b_1,\ldots,a_r,b_r\geq 1}
a_1a_2\ldots a_rC_{a_1b_1}\ldots C_{a_rb_r}
{\partial\over{\partial C_{a_1+\ldots+a_r+n,b_1+\ldots+b_r}}}\0\\
&&+\hbox{ lower orders}\0
\b
and the corresponding ${\cal L}^{[r]}_n(1)$.
All these generators form a closed $W_{1+\infty}$--algebra
\ai
&&\relax
[T^{[r]}_n(1), T^{[s]}_m(1)]=(sn-rm)T^{[r+s-1]}_{n+m}(1)+\ldots,\label{c17a}\\
&&[{\cal L}^{[r]}_n(1), {\cal L}^{[s]}_m(1)]=(sn-rm)
 {\cal L}^{[r+s-1]}_{n+m}(1)+\ldots,\label{c17b}\\
&&[T^{[r]}_n(1),{\cal L}^{[s]}_m(1)]=0,\label{c17c}
\bj
for $r,s\geq1;~n\geq-r,m\geq-s$. Here dots denote lower than $r+s-1$ rank
operators.
Since we have already shown that a combination of the rank--$2$ and
rank--$3$ operators
{\it annihilate} the partition function, eqs.(\ref{c5})
and (\ref{c10}), then, due to the above $W_{1+\infty}$ algebra
\footnote{In fact we have two $W_{1+\infty}$ algebras, one formed by the
$T$ generators and the other by the ${\cal L}$ generators, but since for the
constraints the relevant algebra is the direct sum $T_n+{\cal L}_n$,
we keep speaking about a unique algebra.}, the same
combination of higher rank
generators will annihilate the partition function, that is to say
\a
({\cal L}^{[r]}_n(1)-(-1)^rT^{[r]}_n(1))Z_N(t;C)=0,\qquad r\geq1;\qquad
n\geq-r.
\label{c16}
\b

In the same way from the second linear system we can derive another copy of
$W_{1+\infty}$ algebra as well as another version of the $W_{1+\infty}$
constraints
\a
\bigl({\cal L}^{[r]}_n(2)-(-1)^r T^{[r]}_n(2)\bigl)Z_N(t;C)=0,\qquad r\geq1.
\label{c18}
\b
where ${\cal L}^{[r]}_n(2)$ can be obtained from ${\cal L}^{[r]}_n(1)$
by simply replacing the $t_{1,k}$ parameters with the $t_{2,r}$ ones. However,
as for the $T^{[r]}_n(2)$ generators, we have
\ai
T^{[1]}_n(2)&\equiv&\sum_{a,b\geq 1}bC_{ab}
{\partial\over{\partial C_{a,b+n}}},\qquad n\geq -1\0\\
T^{[2]}_n(2)&\equiv&\sum_{a,b,a^{'},b^{'}\geq 1}bb^{'}C_{ab}C_{a^{'}b^{'}}
{\partial\over{\partial C_{a+a^{'},b+b^{'}+n}}}\0\\
&&+\sum_{a,b\geq 1}b(b-1)C_{ab}
{\partial\over{\partial C_{a,b+n}}},\qquad n\geq -2,\label{c18a}\\
T^{[3]}_n(2)&\equiv&\sum_{a_1,b_1,a_2,b_2,a_3,b_3\geq 1}
b_1b_2b_3C_{a_1b_1}C_{a_2b_2}C_{a_3b_3}
{\partial\over{\partial C_{a_1+a_2+a_3,b_1+b_2+b_3+n}}}\0\\
&&+\sum_{a_1,b_1,a_2,b_2\geq 1}{3\over2}b_1b_2\bigl(b_1+b_2
-2\bigl)C_{a_1b_1}C_{a_2b_2}
{\partial\over{\partial C_{a_1+a_2,b_1+b_2+n}}}\0\\
&&+\sum_{a,b\geq 1}b(b-1)(b-2)C_{ab}
{\partial\over{\partial C_{a,b+n}}},\qquad n\geq -3,\label{c18b}\\
&~&\qquad\qquad \ldots\ldots\0\\
T^{[r]}_n(2)&\equiv&\sum_{a_1,b_1,\ldots,a_r,b_r\geq 1}
b_1b_2\ldots b_rC_{a_1b_1}\ldots C_{a_rb_r}
{\partial\over{\partial C_{a_1+\ldots+a_r,b_1+\ldots+b_r+n}}}\0\\
&~&\qquad\qquad+\hbox{lower order terms},\label{c18c}
\bj
with the following algebra, which is isomorphic to (\ref{c17a}--\ref{c17c})
\ai
&&\relax [T^{[r]}_n(2),
T^{[s]}_m(2)]=(sn-rm)T^{[r+s-1]}_{n+m}(2)+\ldots,\label{c19a}\\
&&[{\cal L}^{[r]}_n(2), {\cal L}^{[s]}_m(2)]=(sn-rm)
 {\cal L}^{[r+s-1]}_{n+m}(2)+\ldots,\label{c19b}\\
&&[T^{[r]}_n(2), {\cal L}^{[s]}_m(2)]=0,\label{c19c}
\bj
for $r,s\geq1;~n\geq-r,m\geq-s$.

In conclusion, we see that, in the case of a general interaction,
there are two isomorphic $W_{1+\infty}$ algebra constraints .
The closed algebra formed by these two pieces of $W_{1+\infty}$
algebras gives the complete constraints (one should notice
that they do not form a direct product).

\subsection{$W_{1+\infty}$ constraints}

{}From eq.(\ref{c16}), it is
easy to see that if we return to our original model, i.e. we set all the
$C_{ab},a,b\geq1$ equal to zero, and only keep $C_{11}=c_{12}=c\neq0$, we have
\ai
&&T^{[r]}_n(1)Z_N(t;C)
=c^r\rtr\Bigl(Q^{r+n}(1)Q^r(2)\Bigl)Z_N(t;C)\label{c20a}\\
&&T^{[r]}_n(2)Z_N(t;C)
=c^r\rtr\Bigl(Q^r(1)Q^{r+n}(2)\Bigl)Z_N(t;C).\label{c20b}
\bj
Substituting these into eqs.(\ref{c16}) and (\ref{c18}), we get
the following constraints
\a
W^{[r]}_nZ_N(t,C)=0,\quad r\geq0;~~n\geq-r,\label{c21a}
\b
where
\a
W^{[r]}_n\equiv (-c)^n{\cal L}^{[r]}_n(1)-{\cal L}^{[r+n]}_{-n}(2).\label{c21b}
\b
form a $W_{1+\infty}$ algebra.
One can explicitly check that this result coincides with the one
in ref.\cite{IM}.

A further reduction is possible. Suppose we set
$t_{2,k}=0, k>q$, then from the above equation, we have
\a
{\cal L}^{[r]}_{-r}(1)+c^r\rtr\bigl(Q^r(2)\bigl)=0
\b
Substituting it into the other constraints, we can get other
$W_{1+\infty}$ constraints, which are only expressed in
terms of $t_{2,k}$ with $1\leq k\leq q$
(beside all the $t_{1,k}$'s, of course). The operators form a
subalgebra of the $W_{1+\infty}$ in eq.(\ref{c21a}--\ref{c21b}).

\section{Differential~~ Hierarchies~~ of~~ Multi--Matrix ~~~~ Models.}

\setcounter{equation}{0}

So far we have shown that multi--matrix models can be represented by means of
coupled discrete linear systems, whose consistency conditions give rise to
discrete KP hierarchies and string equations. In this section, we will
use the method introduced in \cite{BX1} to transform the discrete linear
systems into equivalent differential systems whose consistency conditions are
purely differential hierarchies.
In the most general case they coincide with the generalized \kph.
In this elaboration no continuum limit is involved.

As anticipated in the introduction the clue to
the construction are
the $t_{1,1}$ and $t_{q,1}$ flows.
On the one hand, it is just the first flow equations of $\Psi$ and $\Phi$ that
enable us to eliminate the difference operations from the
RHS of the other flow equations and to
recast the discrete linear systems into a purely differential
form. On the other hand, using them, we can express the flows of the
coordinates of the $j$--th sector
as functions of the coordinates in the same sector. This remarkable
property is far from obvious a priori.

\vskip 0.4cm
\noindent
(*)\underline{\it Differential linear system I.}
\vskip 0.2cm

We apply the procedure outlined above to the discrete linear systems derived
in the previous section. Since for \qmm~ we derived
$2q$ discrete linear systems, when we
pass to the differential language we expect to find $2q$ differential
linear systems.
We start with the DLS (\ref{DLS1}).
Using the eqs.(\ref{rf11}) and (\ref{rfq1}), we have
\a
&&a_0(j-1)=a_0(j)-(\ln R_j)'\0\\
&&a_0(j-2)=a_0(j)-(\ln R_j)'-\bigg(\ln\Bigl[R_j-\dot a_0(j)+
{{\partial^2\ln R_j}\over{\partial t_{1,1}\partial t_{q,1}}}\Bigl]\bigg)'\0
\b
where for any function $f(t)$, we write
\a
f'\equiv {{\partial f}\over{\partial t_{1,1}}}\equiv \partial f,
\qquad\qquad \dot f\equiv  {{\partial f}\over{\partial t_{q,1}}}\equiv \dtl f\0
\b
In general, defining
\a
&&R_{j+r}\equiv F^+_r(j),\qquad a_0(j+r)\equiv G^+_r(j)\0\\
&&R_{j-r}\equiv{F}^-_r(j),\qquad a_0(j-r)\equiv{G}^-_r(j)\0
\b
we obtain the recursion relation
\ai
&&F^+_{r+1}(j)=F^+_r(j)+\dot G^+_r(j)\\
&&G^+_{r+1}(j)=G^+_r(j)+\Bigl(\ln[F^+_r(j)+\dot G^+_r(j)]\Bigl)'\\
&&{F}^-_{r-1}(j)={F}^-_r(j)-\dot{G}^-_r(j)+
{{\partial^2\ln {F}^-_r}\over{\partial t_{1,1}\partial t_{q,1}}}\\
&&{G}^-_{r-1}(j)={G}^-_r(j)-(\ln{F}^-_r)'
\bj
These results guarantee: 1) that  all the $a_0(i)$'s and $R_i$'s ($i\neq j$)
can be expressed in terms of $a_0(j)$ and $R_j$ and their
$t_{1,1}$ and $t_{q,1}$--derivatives; 2) that, substituting these results
into eqs.(\ref{f11}--\ref{falq1}), we can recursively
express all $a_l(i)$'s and $T^{(\al)}_l(i)$'s($i\neq j$) as functions of
the coordinates in the j--th sector.

Now let us consider the $t_{1,1}$--,and $t_{q,1}$--flows of $\Psi_j$.
{}From eqs.(\ref{DLS1}), we can write down their explicit forms
\a
{\partial\over{\partial t_{1,1}}}\Psi_j=\Psi_{j+1}+a_0(j)\Psi_j\label{tpsi11}\\
{\partial\over{\partial t_{q,1}}}\Psi_j=-R_j\Psi_{j-1}\label{tpsiq1}
\b
which lead to the following equalities
\a
\Psi_{j+1}=\hat A_j\Psi_j,\qquad
\Psi_{j-1}=\hat B_{j-1}\Psi_j\label{AB}
\b
where
\a
&&{\hat A}_j\equiv \partial-a_0(j)=-\dtl^{-1}R_{j+1}\0\\
&&{\hat B}_j\equiv \d^{-1}\sum_{l=0}^{\infty}
\bigl(a_0(j)\d^{-1}\bigl)^l=-{1\over{R_{j+1}}}\dtl\0
\b
It is easy to see that
\a
\hat A_j\hat B_j=\hat B_j\hat A_j=1\qquad\forall j\geq1\0
\b
Using eq.(\ref{AB}) we can rewrite the spectral equation in
(\ref{DLS1}) as a {\it purely} differential equation
\a
L_j(1)\Psi_j=\lambda_1\Psi_j\label{spetral1}
\b
where
\a
&&L_j(1)=\d+\sum_{l=1}^{[j,m_1]}a_l(j)\hat B_{j-l}\hat B_{j-l+1}\ldots
\hat B_{j-1}\0\\
&&\qquad=\d+\sum_{l=1}^{[j,m_1]}a_l(j)
{1\over{\d-a_0(j-l)}}\cdot
{1\over{\d-a_0(j-l+1)}}\cdots{1\over{\d-a_0(j-1)}}\0\\
&&\qquad \equiv\d+\sum_{l=1}^{\infty}u_l(j)\d^{-l}
\label{Lj1}
\b
where ${[j,m_1]}$ denotes the least between $j$ and $m_1$.
All the functions $u_l(j)$ are functions of  the coordinates in
j--th sector only.
\a
&&u_1(j)=a_1(j),\0\\
&&u_2(j)=a_2(j)+a_1(j)\bigg(a_0(j)-(\ln R_j)'\bigg)\label{u}\\
&&u_3(j)=a_3(j)+a_2(j)
\bigg[2a_0(j)-2(\ln R_j)'-\bigg(\ln\Bigl[R_j-\dot a_0(j)+
{{\partial^2\ln R_j}\over{\partial t_{1,1}\partial t_{q,1}}}\Bigl]\bigg)'
\bigg]\0\\
&&~~~~~~~~~~+a_1(j)\bigg([a_0(j)-(\ln R_j)']^2-a_0'(j)+
(\ln R_j)''\bigg)\0
\b
and so on. $L_j(1)$ is an operator of KP type. Actually it is in general a
reduction of the KP operator
\a
L_{KP}= \d+\sum_{l=1}^{\infty}w_l(j)\d^{-l}\0
\b
where $w_l$ are generic (i.e. unrestricted) coordinates.
We notice that due to the limitation in the first two summations in
(\ref{Lj1}),
the operators $L_j(1)$ do not have {\it in general} a universal form in terms
of the coordinates $a_l(j)$. In fact they
change with $j$ as long as $j< m_1$, while the form of $L_j(1)$
is independent of $j$ if $j\geq m_1$. We will comment upon this point later on.

Without repeating all the steps of \cite{BX1},
the rules to pass
from lattice to purely differential language are:

$(i)$. The Jacobi matrices are  mapped into KP--type operators, i.e.
\a
\left\{\ba{l}
\Bigl(Q(\al)\Psi\Bigl)_j\Longrightarrow L_j(\al)\Psi_j,\qquad\al
=1,2,\ldots,q\\\noal
\Bigl(P(1)\Psi\Bigl)_j\Longrightarrow M_j(1)\Psi_j.
\ea\right.
\b
where
\ai
L(\al)_j&=&T^{(\al)}_0(j)+\sum_{l=1}^{[j,m_\alpha]}
T^{(\al)}_{-l}(j)\hat A_{j+l-1}\hat A_{j+l-2}\cdots\hat A_j
\0\\
&&+\sum_{l=1}^{[j,n_\alpha]}T^{(\al)}_l(j)\hat B_{j-l}\hat B_{j-l+1}\cdots
\hat B_{j-1}\\
M_j(1)&=&P_0^{(1)}(j)+\sum_{l=1}^{\infty}P_{-l}^{(1)}(j)
\hat A_{j+l-1}\hat A_{j+l-2}\cdots\hat A_j
\0\\
&&+\sum_{l=1}^{[j,s_1]}P^{(1)}_l(j)\hat B_{j-l}\hat B_{j-l+1}\cdots\hat B_{j-1}
\bj
We remark again the non--universality of such expression except for $j$ large
enough.

$(ii)$. The lower triangular part of a Jacobi matrix (or its powers) maps
to the pure integration part of the KP operator (or its powers);

$(iii)$. The upper triangular part together with the main diagonal line of
the Jacobi matrix (or its powers) correspond to the purely differential part
of the KP operator (or its powers).

$(iv)$. The residue of the KP operators have a particularly simple
form,
\a
\res_{\d}(L_j(1))=a_1(j)=Q_{j,j-1}(1).\label{resq}
\b
More generally, we can obtain
\a
\res_{\d}\bigl(L_j^r(1)\bigl)=Q^r_{j,j-1}(1).\label{resqr}
\b
and
\a
\res_{\d}(L_j(\al))=T^{(\al)}_1(j)=Q_{j,j-1}(\al).\0
\b
as well as
\a
\res_{\d}(L^r_j(\al))=\Bigl(Q^r(\al)\Bigl)^r_{j,j-1}(\al).\label{resqal}
\b

Expanding all the above operators in powers of $\d$, the coefficients can
be expressed as functions of the coordinates in the j--th sector.
Collecting all the results we get a linear system
\a
\left\{\ba{l}
L_j(1)\Psi_j=\lm_1\Psi_j\\\noal
\frac {\d}{\d t_{1,r}}\Psi_j=\Bigl(L^r_j(1)\Bigl)_+\Psi_j\\\noal
\frac {\d}{ \d t_{\al,r}}\Psi_j=-\Bigl(L^r_j(\al)\Bigl)_-\Psi_j,\qquad
\al=2,3,\ldots,q\\\noal
M_j(1)\Psi_j=\ddlm 1\Psi_j
\ea\right.\label{diffls1}
\b
The consistency conditions are
\a
&&\frac{\partial}{\partial t_{\beta r}} L_j(\alpha)=
[ \bigl(L_j^r(\beta)\bigr)_+,
L_j(\alpha)],
\qquad\qquad 1\leq \beta\leq \alpha \label{cflowa}\\
&&\frac{\partial}{\partial t_{\beta r}} L_j(\alpha)=[ L_j(\alpha),
\bigl(L_j^r(\beta)\bigr)_-],
\qquad\qquad \alpha \leq \beta\leq n \label{cflowb}
\b
We do not drop here
the label $j$ in the above equations in order to stress the fact that, although
the hierarchy is expressed in terms of the coordinates of the $j$--th sector
only, $L_j$ depends on j, as we have already remarked.

The linear system (\ref{diffls1}) is exactly the same as the one obtained
in \cite{X2} in a completely different context, i.e. by trying to
generalize the KP hierarchy.

We remark that if we impose the condition
\a
a_l(j)=0,\qquad\qquad \forall~l\geq2\0
\b
then the second expression of (\ref{Lj1}) gives the so--called two--bosonic
representation of the KP hierarchy. Motivated by this fact,
we refer to the full expression
of eq.(\ref{Lj1}) as the multi--bosonic representation of the KP hierarchy.

\vskip 0.4cm
\noindent
(**)\underline{\it Differential linear system II.}
\vskip 0.2cm

In the above analysis, we considered $t_{1,1}$ as the space coordinate, and
$\d$ as the basic derivative. However, as we already pointed out,
it is also possible to consider $t_{q,1}$ as the space coordinate and
$\dtl$ as the basic derivative. In this way we would get another linear
system, in which $\Phi$ plays the
role of Baker--Akhiezer function. In the following we discuss this system.
As before we have
\a
\Phi_{j+1}=\hat C_j\Phi_j,\qquad
\Phi_{j-1}=\hat D_{j-1}\Phi_j\label{ABprime}
\b
where
\a
&&\hat C_j\equiv \dtl-b_0(j)=-\d^{-1}R_{j+1}\0\\
&&\hat D_j\equiv \dtl^{-1}\sum_{l=0}^{\infty}
\bigl(b_0(j)\dtl^{-1}\bigl)^l=-{1\over{R_{j+1}}}\d\0
\b
The linear system is
\ai
&&\TL_j(q)\Phi_j=\lm_q\Phi_j\\
&&\frac {\d}{\d t_{q,r}}\Phi_j=\Bigl(\TL^r_j(q)\Bigl)_+\Phi_j\\
&&\frac {\d}{\d t_{\al,r}}\Phi_j=-\Bigl(\TL^r_j(\al)\Bigl)_-\Phi_j,\qquad
\al=1,2,\ldots,q-1\\
&&\TM_j\Phi_j=\ddlm q\Phi_j
\bj
Now the KP operator takes the following form
\a
&&\TL_j(q)=\dtl+\sum_{l=1}^jb_l(j)\hat D_{j-l}\hat D_{j-l+1}\ldots
\hat D_{j-1}\0\\
&&\qquad=\dtl+\sum_{l=1}^jb_l(j)
{1\over{\dtl-b_0(j-l)}}
{1\over{\dtl-b_0(j-l+1)}}\ldots{1\over{\dtl-b_0(j-1)}}\0\\
&&\qquad \equiv\dtl+\sum_{l=1}^{\infty}v_l(j)\dtl^{-l}
\label{TLjq}
\b
All the functions $v_l(j)$ are only functions of  the coordinates in
the $j$--th sector.
\a
&&v_1(j)=b_1(j),\0\\
&&v_2(j)=b_2(j)+b_1(j)\bigg(b_0(j)-\dtl(\ln R_j)\bigg)\label{v}\\
&&v_3(j)=b_3(j)+b_2(j)\bigg[2b_0(j)-2\dtl(\ln R_j)
-\dtl\bigg(\ln\Bigl[R_j-\d b_0(j)+
{{\partial^2\ln R_j}\over{\partial t_{1,1}\partial t_{q,1}}}\Bigl]\bigg)
\bigg]\0\\
&&~~~~~~~~~+b_1(j)\bigg([b_0(j)-\dtl(\ln R_j)]^2-\dtl {b}_0(j)+
\dtl ^2{\ln R_j}\bigg)\0
\b
and so on. Using the rules to go from the lattice to the differential
language, we obtain
\ai
&&\Bigl(\q(\al)\Phi\Bigl)_j\Longrightarrow \TL_j(\al)\Phi_j,
\qquad\al=1,2,\ldots,q\\
&&\Bigl(P(q)\Phi\Bigl)_j\Longrightarrow {\tilde M}_j(q)\Phi_j.
\bj
with new differential operators
\ai
&& \TL (\al)_j=T^{(\al)}_0(j)+\sum_{l=1}^{\infty}T^{(\al)}_{-l}(j)
\hat C_{j+l-1}
\hat C_{j+l-2}\cdots\hat C_j\0\\
&&\qquad\quad+\sum_{l=1}^jT^{(\al)}_l(j)
\hat D_{j-l}\hat D_{j-l+1}\cdots\hat D_{j-1}\\
&&\TM_j(q)=P_0^{(q)}(j)+\sum_{l=1}^\infty P_{-l}^{(q)}(j)
\hat C_{j+l-1}\hat C_{j+l-2}
\cdots\hat C_j
\0\\
&&\qquad\quad+\sum_{l=1}^{s_q}P_l^{(q)}(j)\hat D_{j-l}\hat D_{j-l+1}
\cdots\hat D_{j-1}
\bj
Therefore we can write down another linear system
\a
\left\{\ba{l}
\TL_j(q)\Phi_j=\lm_q\Phi_j\\\noal
{{\d}\over {\d t_{q,r}}}\Phi_j=\Bigl(\TL^r_j(q)\Bigl)_+\Phi_j\\\noal
{{\d}\over {\d t_{\al,r}}}\Phi_j=-\Bigl(\TL^r_j(\al)\Bigl)_-\Phi_j,\qquad
\al=1,2,\ldots,q-1\\\noal
M_j(1)\Phi_j=\ddlm 1\Phi_j
\ea\right.\label{diffls2}
\b
whose consistency conditions also give the hierarchy
(\ref{cflowa}--\ref{cflowb}).

We could also start from the
general form of the discrete linear systems (\ref{DLS1al}) and
(\ref{DLS2al}), translate them into the differential language by  using
the rules we listed before.
Once again they lead to the hierarchy (\ref{cflowa}--\ref{cflowb}).
We will spare the reader the details.

We end this section with a remark on
the flows in (\ref{cflowa}) and (\ref{cflowb}). They all
commute with one another, since their lattice versions do.
So all the differentiable hierarchies (\ref{cflowa}) and (\ref{cflowb})
are integrable.
This is a very remarkable result. We know in fact that attempts at enlarging
the KP hierarchy in a very general way have led to incompatible
KP flows\cite{Dickey}.
Therefore we may say that the multi--matrix models give a natural realization
of commutating KP flows.

\subsection{The coupling conditions}

Our final task is to reexpress the coupling conditions
(\ref{coup1}--\ref{coup2}) in the differential language. This can be done
very easily by replacing matrices by their operator versions. The only
problem we should care about is the matrix $\PBC$, since its differential
version is not $\TM$(q). We may simply denote its differential version
by $M(q)$, then the coupling conditions are
\ai
&&M(1)+c_{1,2}L(2)=0,\label{ml2}\\
&&c_{\alpha-1,\alpha}L(\alpha-1)+V'(\alpha)+c_{\alpha,\alpha+1}
L(\alpha+1)=0,\qquad 2\leq\al\leq q-1;\label{loperator}\\
&&c_{q-1,q}L(q-1)+M(q)=0.\label{tmlq-1}
\bj
where for simplicity we dropped the label $j$, denoting the sector.
Of course
\a
V'(\alpha)= \sum_{r=1}^{p_\alpha}r t_{\alpha,r}L^{r-1}(\alpha)\0
\b

Some of these equations are also met in  \cite{X2}, so
they are characteristic of the generalized KP hierarchy.
But not all of them appear simultaneously in the generalized KP hierarchy.
More precisely, if we are in the $t_{1,1}$--picture, that is $t_{1,1}$ is
assumed to be the `space' variable, then in the generalized KP hierarchy
there naturally appear only (\ref{ml2}) and (\ref{loperator}); therefore
in this picture (\ref{tmlq-1}) is a constraint on the generalized KP hierarchy.
Vice, in the $t_{q,1}$--picture (\ref{tmlq-1}) and (\ref{loperator})
appear naturally in the generalized KP hierarchy, while (\ref{ml2}) plays the
role of constraint.

Thus we may conclude that multi--matrix models correspond
to the generalized KP hierarchy subject to certain constraints.

\subsection{The partition function as $\tau$--function}

In the generalized KP hierarchy we can define the following $\tau$--function
\a
\frac{\partial^2}{\partial t_{1,1} \partial t_{\alpha,r}} \ln \tau=
\res_\partial L^r(\alpha), \quad\quad \forall~~\alpha,r\label{taumulti}
\b
where $L$ is the generalized $KP$ operator.
We want to prove that the partition function of \mmm~  is a
$\tau$--function of the $\alpha$-th generalized KP hierarchy.
To achieve this we only need to prove
that eqs.(\ref{parti3}), once translated into the differential language,
have the same form as eqs.(\ref{taumulti}).
This can be easily done, if we note that
eq.(\ref{resqal}) is valid
for any $1\leq\al\leq q$, and
any positive integers $r,j$. Choosing in particular $j=N$, we have
\a
\Bigl(Q^r(\al)\Bigl)_{N,N-1}
=\res_{\d}\Bigl(L_N(\al)\Bigl)^r
\b
This equality together with eq.(\ref{parti3}) gives
\a
{\d^2\over{\d t_{1,1}\d t_{\al,r}}}\ln Z_N(t,c)
=\res_{\d}\Bigl(L_N(\al)\Bigl)^r
\b
This is nothing but (\ref{taumulti}).
Therefore we can conclude that the partition functions of multi--matrix
models are exactly $\tau$--functions of the generalized KP hierarchy.

\section{Examples and Conclusion}

\setcounter{equation}{0}

In the previous sections we were able to extract differential hierarchies
from \mmm, without passing through a continuum limit but by simply studying
the properties of the matrix model lattice. These are KP type hierarchies
of which the original partition function is a $\tau$--function. To determine
the initial conditions for this $\tau$--function we digged out additional
equations: the coupling conditions or string equations.

What we have worked out so far is a very general scheme which allows for
a vast number of possibilities, i.e. physical models, the physical meaning
being attached to the coordinates we have introduced.
The analysis of all the different
possibilities is out of the question. Our present purpose is much more modest:
we want to single out significant subclasses of models. To do this we can play
with two tools: the first is the type of potentials introduced in the
original multi--matrix path--integral; the second tool, once the potentials
are fixed and the corresponding differential linear systems are worked out,
is to study the various possible reductions of the latter.

To start with we want to split all the possibilities we are faced
with in two broad classes or alternatives.
We recall what we repeatedly pointed out in the last section:
the differential hierarchies obtained from the lattice are in general
not universal in that they depend on $j$, i.e. on the sector. In particular,
the flows
of the logarithm of the partition function depend in general on the sector.
For the differential hierarchies coming from a given lattice hierarchy
we have therefore two possibilities, which are connected with specific
choices of the numbers $p_\alpha$ characterizing the order of the potential
$V_\alpha$:

-- the {\it non--universal alternative}: the differential hierarchies coming
from the same lattice hierarchy are all different in different sectors.
In this case one can hardly consider these hierarchies an intrinsic
property of the lattice. Presumably the only way (if any) to treat this case is
via a continuum limit.

-- the {\it universal alternative}: the differential hierarchies coming
from the same lattice hierarchy become isomorphic, i.e. truly independent
of $j$, for $j$ large enough. No doubt, the $j$--independent hierarchies,
referred to henceforth as the relevant {\it universal hierarchy},
express intrinsic properties of the lattice, and they can presumably be
realized in terms of topological field theories. Needless to say, we
are interested in this case.

Let us see, as an example, the two--matrix model case.
We have two parameters: $p_1$ and $p_2$. On the basis of the discussion
in subsection 3.1, we see that the lattice {\it system I} gives rise
to a universal hierarchy only if $p_2$ is finite, while the lattice {\it system
II} does the same only if $p_1$ is finite. Generalizing this discussion to
other \mmm~ is straightforward.

This distinction between universal and non--universal alternative is perhaps
difficult to grasp at first reading. However it has a
rather natural meaning: we recall once again that the
information concerning the multi--matrix model is contained in the partition
function; the flows of the latter depend on the sector $j=N$ (see subsection
5.2); in the universal alternative the potentials are simple enough as to allow
us to retrieve all the information contained in the lattice if we choose
a sufficiently large (but finite) $N$, while of course only partial
information is retrieved if $N$ is too small; in the non--universal case
the potentials are too complicated for us to retrieve all the lattice
information with a finite $N$, but we might succeed perhaps
when $N\rightarrow \infty$.\footnote{We advise the reader not to rely too much
in this regard on the
analogy with \omms. The analogy between \omms~ and \mmm~ is only partial.
One--matrix models are all to be classified in the universal alternative, as
their Jacobi matrix has at most three non--vanishing diagonal lines. The
dependence of equations and differential operators on the sector is instead
present in \omms~ as well, although in a quite unconspicuous way as it appears
only in the zeroth sector while
everything takes the universal form starting from the first sector on.}

To end this paper we would like to give a few more explicit examples of the
differential systems of the last section. We will consider the differential
{\it system I} in multi--matrix models; not in full generality however,
only in the dispersionless limit \cite{Kr}, \cite{Du}. This limit is in any
case very interesting.
It represents the genus 0 approximation and it might characterize the full
hierarchy. So let us concentrate on the {\it system I} of the previous
section and,
in particular, on the differential operator $L_j(1)$, eq.(\ref{Lj1}), with
the following specifications: $p_1=\infty$ and $p_2$ finite,
so that $L_j(1)$ satisfies a universal hierarchy for large enough $j$. For this
reason we will drop in the following the label $j$.

The technique to obtain the dispersionless hierarchy is very simple once we
know the full hierarchy: in the flow equations we simply discard terms
containing more than one first order derivative or higher order derivatives.
A simple way to obtain it
without previous knowledge is to replace $\partial$ with a variable $p$ which
is
Poisson conjugate to $t_1$,
\a
\{p, t_1\}=1\0
\b
and fully exploit the integrable structure. Therefore we start with the
replacement
\a
L(1)\Longrightarrow {\cal L}= p+\sum_{l=1}^\nu a_l \frac{1}{(p-S)^l}
\equiv p+\sum_{l=1}^\infty u_l ~p^l\label{dispL}
\b
where
\a
u_{l+1}= \sum_{k=0}^l \left(\matrix{l\cr k\cr}\right)a_{k+1} S^{l-k},
\quad\quad\quad a_l=0~~{\rm for}~~l>\nu\0
\b
In (\ref{dispL}) we set $a_0=S$ in analogy with \cite{BX1}.
At this point, in order to find the flows of these coordinates, we could
simply use the consistency conditions of the {\it system I} where commutators
are replaced by corresponding Poisson brackets. A better way is to use the
Poisson algebra. The first Poisson bracket is given by
\a
&&\{a_1(x), S(y)\}_1= \delta'(x-y)\label{1P}\\
&&\{a_i(x), a_j(y)\}_1= \Big[(i+j-2)a_{i+j-2}\partial +(j-1)a_{i+j-2}'\Big]
\delta(x-y),\quad\quad 2\leq i,j\leq \nu\0
\b
while all the other brackets vanish.
The second Poisson bracket is
\a
\{S(x),S(y)\}_2&=&\textstyle\frac{\nu+1}{\nu} \delta'(x-y)\label{2P}\\
\{S(x), a_1(y)\}_2&=& \partial S\delta(x-y)\0\\
\{S(x), a_j(y)\}_2&=&\textstyle \frac{\nu-j+1}{\nu} \partial a_{j-1}
\delta(x-y)\0\\
\{a_1(x), a_1(y)\}_2&=& \Big[a_1\partial +\partial a_1\Big] \delta(x-y)\0\\
\{a_1(x), a_j(y)\}_2&=& \Big[(j+1)a_j\partial + j a_j'\Big] \delta(x-y)\0\\
\{a_i(x), a_j(y)\}_2&=& \Big[\Bigl(ia_{i+j-1} \partial+ j \partial a_{i+j-2}
+(i-1)a_{i+j-2}\partial S\0\\
&&+(j-1) S \partial a_{i+j-2}
-(i-1)\textstyle\frac{\nu-j+1}{\nu} a_{i-1}\partial a_{j-1}
\Bigr)\partial\0\\
&&+\sum_{l=1}^{i-2} \Bigl((i-l-1)a_{i+j-l-2} \partial a_l +(j-l-1) a_l \partial
a_{i+j-l-2}\Bigr)\Big]\delta(x-y)\0
\b
In these equations $i,j\geq 2$.
Next we write down the Hamiltonians
\a
{\cal H}_r = \frac{1}{r} {\cal L}^r_{(-1)}\label{dispH}
\b
The RHS denotes the coefficient of $p^{-1}$ in ${\cal L}^r$. The two Poisson
brackets are compatible with respect to these Hamiltonians. Using this fact
we can easily write down the flows in terms of
\a
F_r= \frac{\delta {\cal H}_r}{\delta S}, \quad\quad
G_r= \frac{\delta {\cal H}_r}{\delta a_1}, \quad\quad
E_r= \frac{\delta {\cal H}_r}{\delta a_2}, \quad\quad\ldots\0
\b
and recursion relations for the latter. For example, in the case of
three fields, i.e. $\nu=2$, we obtain
\a
&&\frac{\partial S}{\partial t_r} = G'_{r+1}\0\\
&&\frac{\partial a_1}{\partial t_r} = F'_{r+1}\label{dispflow}\\
&&\frac{\partial a_2}{\partial t_r} = 2a_2E'_{r+1}+a_2'E_{r+1}\0
\b
The recursion relations are
\a
&&G'_{r+1}= \frac{3}{2} F'_r + (SG_r)' + \frac{1}{2} (a_1E_r)'\0\\
&&F'_{r+1}= SF'_r + 2a_1 G'_r +a_1'G_r +3a_2E'_r + 2a_2' E_r\label{disprec}\\
&&2a_2E'_{r+1}+ a_2'E_{r+1}= \frac{1}{2} a_1 F'_r+3a_2 G'_r +a_2'G_r\0\\
&&~~~~~~~~~~~~~~~~~~~~~~~~~~~+(2a_2S-\frac{1}{2} a_1^2)E'_r + (a_2 S-
\frac{1}{4}a_1^2)'E_r\0
\b
the initial conditions being $F_1=E_1=G_1=0$.

{}From this result we can learn something important. In \cite{BX1},\cite{BX2}
we obtained the KdV hierarchy reducing the NLS hierarchy. The latter is
generated as the consistency condition of {\it system I} ~for a KP operator
of the type $L(1)$ with only two non--vanishing coordinates $a_0=R$ and $S$.
The reduction to the KdV hierarchy is obtained by setting $S=0$. Therefore
it would seem not unmotivated to expect that setting $S=0$ in (\ref{dispflow})
one finds the dispersionless 3--KdV (or Boussinesq) hierarchy. However
an explicit calculation proves that this is not the case. In other words this
approach to find the n--KdV hierarchies as reductions of
the differential systems ensuing from \mmm~ is too naive.
A way to obtain the n--KdV hierarchy could be as follows:
reduce $L(1)$ in just the same way
we reduce the KP operator down to the n--KdV operator; if the order of the
$V_2$ potential (in two--matrix model) is large enough, we can obtain
all the flows we wish of the n--KdV hierarchy.

In any case it is evident that the reduction to the n--KdV hierarchies is
a very drastic one: most of the generalized KP structure gets lost in this
passage. Taught by the \omm~ example, \cite{BX2}, we think a lot of
information,
in other words of topological models, is contained in the generalized KP
structure as it appears in \mmm. We intend to deal elsewhere with this part
of the analysis of \mmm.

\end{document}